# Anomalous optical phonons in FeTe pnictides: spin state, magnetic order, and lattice anharmonicity

V. Gnezdilov,[1] Yu. Pashkevich,[2] P. Lemmens,[3] A. Gusev,[2] K. Lamonova,[2] I. Vitebskiy,[4] O. Afanasiev,[1] S. Gnatchenko,[1] V. Tsurkan,[5,6] J. Deisenhofer,[5] and A. Loidl[5]

[1]*B. I. Verkin Institute for Low Temperature Physics and Engineering of the National Academy of Sciences of Ukraine 47 Lenin Ave., Kharkov 61103, Ukraine,*

[2]*A. A. Galkin Donetsk Phystech of the National Academy of Sciences of Ukraine, Donetsk 83114, Ukraine,*

[3]*Institute for Condensed Matter Physics, TU Braunschweig, D-38106 Braunschweig, Germany,*

[4]*University of California at Irvine, Irvine, CA 92697, USA,*

[5]*Experimental Physics V, Univ. Augsburg, 86159 Augsburg, Germany*

[6]*Institute of Applied Physics, Academy of Sciences of Moldova, MD-2028 Chisinau, R. Moldova*

**Abstract**

Polarized Raman-scattering spectra of non-superconducting, single-crystalline FeTe are investigated as function of temperature. We have found a relation between the magnitude of ordered magnetic moments and the linewidth of $A_{1g}$ phonons at low temperatures. This relation is attributed to the intermediate spin state (S=1) and the orbital degeneracy of the Fe ions. Spin-phonon coupling constants have been estimated based on microscopic modeling using density-functional theory and analysis of the local spin density. Our observations show the importance of orbital degrees of freedom for the Fe-based superconductors with large ordered magnetic moments, while small magnetic moment of Fe ions in some iron pnictides reflects the low spin state of Fe ions in those systems.

# I. INTRODUCTION

The report of superconductivity at 26 $K$ in iron pnictides [1] and iron chalcogenides [2] has triggered an intense burst of research activities comparable to that during the early days after the discovery of superconducting cuprates or hydrated cobaltates $Na_xCoO_2 \cdot yH_2O$. The search for new superconducting materials and the attempts to raise superconducting transition temperatures, $T_c$, by chemical doping [3, 4, 5, 6] and by external pressure [7, 8, 9, 10] have led to the discovery of other members of iron pnictide and chalcogenide families with higher $T_c$. The presence of a layered crystal structure with Fe ions in tetrahedral coordination is a general structural feature of the iron based compounds. This main building block possesses a tetragonal planar symmetry at room temperature, which is associated with certain degeneracy of electronic and phononic spectra. Simultaneously many theoretical scenarios were evaluated, however, not only the superconductivity mechanism but also the magnetic properties of the iron-based compounds remain disputable until now. All scenarios agree with respect to the crucial importance of the distance between the Fe surrounding ligands and the Fe-plane on the electronic ground state. Raman scattering on phonons that modulate these distances could shed light upon the interplay between lattice, charge, orbital, spin degrees of freedom and superconductivity.

The pnictide family $Fe_{1+y}Se_xTe_{1-x}$ occupies a special place among newly discovered iron superconductors. Firstly, the members of this family have very simple stoichiometry, while their crystal structure can be seen as a stack of $FeSe_xTe_{1-x}$ layers. Secondly, $Fe_{1+y}Te$ has an unusual magnetic translation symmetry with an in-plane magnetic propagation vector $k_1 = (\pi/a, 0)$, rather than $k_0 = (\pi/a, \pi/a)$ observed in $Re$FeAsO [11]. This kind of magnetic ordering is associated with the rare phenomenon of (orthorhombic) magnetostriction of purely exchange nature. Finally, the most remarkable feature of $Fe_{1+y}Se_xTe_{1-x}$ is the large magnetic moment, which is the highest among the pnictides reaching 2.5 $\mu_B$/Fe for $Fe_{1.05}Te$ [12] . By comparison, the maximum moment for 1111 materials does not exceed 0.4 $\mu_B$/Fe, and it does not exceed 1 $\mu_B$/Fe for 122 materials [13]. The fact that the intermediate spin state ($S = 1$) of Fe ions is realized in FeTe implies that the orbital degrees of freedom of Fe ions play important role in this compound [14, 15].

In this paper we present the results of theoretical and experimental study of phonon Raman scattering in non-superconducting $Fe_{1+x}Te$. The zone-centered and the Raman-active phonons are classified by the irreducible representations of the space symmetry group of the crystal. First-principles lattice-dynamics calculations are performed for the monoclinic magnetic phase of FeTe. Our theoretical and experimental results appear to be in good agreement. The remarkable temperature dependence of the phonon modes in FeTe is discussed in the context of its electronic properties.

## II. EXPERIMENT

Single crystals of $Fe_{1.051}Te$ were grown using Bridgman and self-flux methods. The actual composition was determined by x-ray analysis as $Fe_{1.051}Te$ (with $a$ = 3.8220(1) Å and $c$ = 6.2889(1) Å). A drastic drop in $\chi(T)$ observed at $T_N \approx 70\ K$ is attributed to antiferromagnetic ordering. Raman scattering experiments were carried out in a quasi-backscattering geometry. A solid-state laser was used for an excitation at 532.1 $nm$. To protect the sample from heating, the laser output power was kept below 5 $mW$ on a focus of approximately 50 $\mu m$ of diameter. The spectra were measured in two polarization configurations (parallel, $XX$ and crossed, $XY$) within the crystallographic ab-plane. The scattered light was collected and dispersed by a triple monochromator DILOR XY on a liquid-nitrogen-cooled CCD detector. The measurements were taken in a variable temperature closed-cycle cryostat (Oxford/Cryomech Optistat, $T = 2.8\ K - 300\ K$).

## III. RESULTS AND DISCUSSION

At room temperature, the single-phase FeTe(Se) has the tetragonal PbO structure (space group $P4/nmm$) [2, 16, 17, 18]. In this phase, iron chalcogenide forms the same edge-sharing antifluorite layers, also found in the FeAs superconductors. Fe and Te ions occupy 2a and 2c Wyckoff positions, respectively. Symmetry analysis shows that there are four Raman-active modes ($A_{1g}$(Te) + $B_{1g}$(Fe) + $2E_g$(Te, Fe)) and two infrared-active modes ($A_{2u}$(Te, Fe) + $E_u$(Te, Fe)). The Raman tensors take the form:

$$A_{1g} = \begin{pmatrix} a & 0 & 0 \\ 0 & a & 0 \\ 0 & 0 & b \end{pmatrix},\quad B_{1g} = \begin{pmatrix} c & 0 & 0 \\ 0 & -c & 0 \\ 0 & 0 & 0 \end{pmatrix},\ \text{and}\quad E_g = \begin{pmatrix} 0 & 0 & -e \\ 0 & 0 & e \\ -e & e & 0 \end{pmatrix}.$$

At $T_s = 70\ K$, the metallic FeTe undergoes a first-order phase transition from tetragonal to monoclinic phase, and below 70 $K$ its space symmetry group is $P2_1/m$ [12, 19, 20]. This phase transition is accompanied by $u_{xz}$ – type distortion and simultaneous antiferromagnetic ordering. Noticeably, the symmetry of the long-range magnetic order below 70 $K$ is such that it is compatible with the observed monoclinic structural distortion accompanying the phase transition. This suggests that, at least from symmetry point of view, crystal distortions below 70 $K$ could be a result of the antiferromagnetic ordering – similar to the phenomenon of magnetostriction. If this is the case, the phase transition at 70 $K$ is magnetic in nature, while the crystal distortions are a secondary effect. A way to prove or disprove this assumption is to destroy the antiferromagnetic order by applying a sufficiently strong magnetic field and see if the structural distortions persist. Since we did not conduct such measurements, we cannot speculate on the physical nature of the phase transition.

In addition to the change in the lengths of the $a$ and $b$ axes, the length of the $c$ axis increases by nearly 0.02 $Å$ and its direction rotates towards the $a$ axis creating a slightly acute angle $\beta = 89.2^o$. As a result, the only remaining 2-fold symmetry axis is the $b$-axis [12, 19]. The $u_{xz}$ distortion leads to the change in the $x$ and $z$ – coordinates for both Fe and Te atoms and to a small corrugation of the Fe-plane. In this case all atoms occupy the same 2e Wyckoff positions each of which contributes to three Raman active modes ($2A_g + B_g$). Raman tensors take the form:

$$A_g = \begin{pmatrix} a & d & 0 \\ d & b & 0 \\ 0 & 0 & c \end{pmatrix}, \text{ and } B_g = \begin{pmatrix} 0 & 0 & e \\ 0 & 0 & f \\ e & f & 0 \end{pmatrix}.$$

The two different 1D representations $A_{1g}$ and $B_{1g}$ of the tetragonal symmetry group correspond to the same 1D representation $A_g$ of the monoclinic symmetry group, while the double degenerate irreducible representation $E_g$ splits into $A_g$ and $B_g$. The latter could lead to some leakage of previously $E_g$ modes to our geometry of Raman spectra measurements at low temperatures. Note that the extra Fe ions in Fe$_{1+y}$Te occupy the same 2c positions as the Te ions in the tetragonal phase and 2e position in the monoclinic phase [12, 19].

The zone-center phonons and the electronic structure of the FeTe in the monoclinic magnetic phase have been calculated within the framework of density-functional theory (DFT). We applied the all-electron full-potential linearized augmented-plane-wave method (ELK code) [21] with the local spin density approximation (LSDA) [22] for the exchange-correlation potential and with the revised generalized gradient approximation of Perdew-Burke-Ernzerhof (PBEsol) [23]. We used the experimental unit-cell parameters from $Fe_{1.05}Te$ at $T = 2$ K [12] without structure optimization. The magnetic unit cell in our model is composed of two crystallographic monoclinic unit cells related by the primitive translation along the *a*-axis. For simplicity, we assume in our computations that the magnetic moments of iron are parallel to the *c*-axis with the same antiferromagnetic sign alternation as that of the real magnetic structure [12]. We did not account for the unit cell doubling along the *c*-axis. Our analysis shows a strong dependence of the phonon frequencies on the magnitude of iron magnetic moments. In our theoretical calculations, the magnitude of iron magnetic moments was chosen to provide the best agreement between theoretical and experimental phononic spectra.

In Fig. 1 shows the polarized Raman spectra of single crystal $Fe_{1.05}Te$ at $T = 290$ and 20 K, in *XX* and *XY* scattering configurations. Two strong lines can be easily identified in the spectra. At room temperature these lines are located at 151 and 197 cm$^{-1}$ and were earlier assigned to $A_{1g}$ (Te) and $B_{1g}$ (Fe) phonon modes, respectively [24, 25, 26]. In Table I we compare our experimental results with numerical simulations. The comparison shows a very good agreement between our theoretical and experimental phononic spectra, provided that we set the iron magnetic moment to be equal to 2.2 $\mu_B$. Surprisingly, this above value turns out to be very close to the experimental value 2.52 $\mu_B$ of the iron magnetic moment [12]. This suggests that for the FeTe the above feature is in accordance with Yildirim's finding for Fe-As 1111 and 122 compounds [27]. Indeed, the latter study also shows a strong relation of the phonon spectra with the magnetic moment of Fe sublattices. Importantly, phonon frequencies obtained in our numerical computations are in much better agreement with experiments than those obtained in earlier modeling [24-26]. The reason for that is that in the computations conducted in [24-26], the magnitude of iron magnetic moment has not been taken into account explicitly. In our analysis, we did not account for the contribution from the excess iron, although, in accordance with scenario developed in Ref. [28], even a small Fe excess strongly changes the FeTe Fermi surface towards a nesting condition along ($\pi$/a, 0).

In Fig. 2 (a-f) we show the temperature dependent parameters of the $A_{1g}$ and $B_{1g}$ phonon modes. The $T$ dependencies of the frequency of both $A_{1g}$ and $B_{1g}$ modes show anomalies in the region of the structural transition and magnetic ordering temperature. Using data from Ref. 12 for the temperature change of the lattices constants of Fe$_{1.05}$Te we fit the frequency temperature dependencies with a Grüneisen law:

$$\frac{\Delta\omega_i(T)}{\omega_i(290)} = -\gamma_i \frac{\Delta V_0(T)}{V_0(290)} \tag{1}$$

Here $V_0$ is the primitive cell volume and all differences are calculated from values at ambient temperature. In the tetragonal phase the Grüneisen parameters equal to $\gamma_{B_{1g}}(Fe) = 2.15$ and $\gamma_{A_{1g}}(Te) = 2.85$ both of which deviate from the usual value $\gamma \approx 2$. The larger deviation for the Te mode evidences a stronger impact of anharmonicity which is expected from the position of this atom in the lattice. At the same time the change of the primitive cell volume $V_0$ in the monoclinic phase demonstrates a rather smooth dependence (Fig. 11 in Ref. 12) Thus, we conclude that the frequency deviation of a phonon from the smoothly varying behavior vs. $T$ may be associated with the onset of magnetic ordering.

The renormalization of the phonon frequency below the magnetic ordering temperature is caused by a phonon modulation of the magnetic interactions which includes superexchange, direct exchange (in metal) and anisotropy [29]. Spin-phonon coupling - $H_{sp}$ presents some quadratic form of the magnetic and the mediated ligand ion's displacements. Supposing that the main contribution to spin-phonon coupling results from the Fe-Te-Fe bond-angle modulation of the exchange interactions we obtain:

$$H_{sp} = \left\{ \lambda_{nnn}^{(I)} \left[ Q_{Fe}^2 + Q_{Te}^2 \right] + \lambda_{nnn}^{(II)} Q_{Fe} Q_{Te} + \lambda_{nn} \left[ Q_{Te}^2 - Q_{Fe}^2 \right] \right\} L_l^2(k_I). \tag{2}$$

Here, $Q_{Fe} = u_{1z}(Fe) - u_{2z}(Fe)$ and $Q_{Fe} = u_{1z}(Te) - u_{2z}(Te)$ are symmetry adapted modes of atom displacements along the $z$-axis for $B_{1g}$(Fe) and $A_{1g}$(Te) phonons, respectively. Atom enumeration and the magnetic structure is shown in Fig. 3. $L_1(k_I) = S_1(k_I) + S_2(k_I)$ – is the magnetic order parameter constructed from Fourier components of the α – sublattice's magnetic moments $S_\alpha(k_I)$. The first and the second term in $H_{sp}$ originate from modulations of the nnn - next nearest neighbors interaction $J_{2a}$ and $J_{2b}$ (here and below we use notation of Ref. 28). The third term comes from the modulation of the nn - nearest neighbor $J_{1a}$ and $J_{1b}$ interactions.

In spite of their exchange origin, the spin-phonon coupling constants $\lambda_{nn/nnn}$ are rather small due to their proportionality to the magnitude of monoclinic distortion $u_{xz}$ which is induced by the first-order structural phase transitions $P4/nmm \rightarrow P2_1/m$. Using this smallness one can derive a magnetically induced frequency shift of $\omega_{Fe}$ and $\omega_{Te}$ phonon frequencies taken at 80 $K$ slightly above $T_N$:

$$\Delta\omega_{Fe/Te} = \frac{1}{2\omega_{Fe/Te}} \left\{ \frac{\lambda_{nnn}^{(I)} - / + \lambda_{nn}}{m_{Fe/Te}} + / - \frac{\lambda_{nnn}^{(II)2}}{(\omega_{Fe}^2 - \omega_{Te}^2)m_{Fe}m_{Te}} \right\} L_1^2(k_I), \qquad (3)$$

here $m_{Fe/Te}$ are atomic masses. From (3) we obtain a renormalization of the phonon frequency in magnetic compounds [30, 31] $\omega_{ph} = \omega_{0ph} + \lambda_{ph}\eta^2$ which traces the temperature dependence of the square of a magnetic order parameter $\eta(T)$. The coupling coefficient $\lambda_{ph}$ is different for each phonon and may have either sign. In our case one can neglect contribution from $\lambda_{nnn}^{(II)}$, which always leads to a repulsion of phonon frequencies, due to its second order smallness. Contributions from the $\lambda_{nnn}^{(I)}$ –term change the phonon frequencies in identical ways either positive or negative. While the contribution from the $\lambda_{nn}$ –term changes the phonon frequencies in the opposite way.

As follows from data shown in Fig. 2 (a, d) at lowest temperature (3 $K$) the shift of the former $B_{1g}$ (Fe) phonon mode is negative while for the former $A_{1g}$ (Te) phonon mode it is positive. Taking into account both these shifts calculated relatively to fitting curves at 3 $K$ we estimated the spin-phonon coupling constants which are $\lambda_{nn} = 3.6(9) \cdot 10^{-6}$ $meV/(Å \mu_B)^2$ and $\lambda_{nnn}^{(I)} = -1.9(5) \cdot 10^{-6}$ $meV/(Å \mu_B)^2$. Here we choose $L_{1y}(k_I) = 5.08$ $\mu_B$ in accordance with data from Ref. 12.

There is another nontrivial contribution to the phonon frequency renormalization which is specific to the magnetic phase in iron pnictides and chalcogenides. Unlike the case of the regular magnetic materials, the antiphase motion of chalcogens (or pnictogens) surrounding Fe ions not only modulates the exchange interactions, but also affects the magnitude of the iron magnetic moment. Here we refer to a recent paper by C.-V. Moon and H. J. Choi [32], where it is stressed that the ordered magnetic moment in FeTe depends on the value of tellurium z–coordinate. This mechanism cannot be reduced to the form given in eqn. (2) or (3) and at lowest temperatures its contribution should be negligible since the magnetic moment becomes temperature independent. Also this mechanism should not be relevant to the Fe $B_{1g}$ vibration as it modulates the Fe-Te distances in an opposite way making them shorter-longer, simultaneously. We therefore relate the unusual frequency shift of the $A_g$ (Te) mode in the magnetically ordered phase to contributions of this mechanism.

A very unexpected behavior of the $A_{1g}$ and $B_{1g}$ phonon linewidth (Fig. 2 (b, e)) is observed in our experiments on $Fe_{1.051}Te$ single crystals. In addition to the anomalies at $T_s$, the FWHM increases pronouncedly on cooling below approximately 150 K. Interestingly, as was found in neutron powder diffraction experiments [33], the FWHM of the (200) peak increases on cooling below the same temperature for all the specimens of $FeSe_{1-x}Te_x$. However, the last observation was interpreted as a decrease in the symmetry of the high-temperature tetragonal structure at around $T = 150$ K even if the (200) peak does not split. The most puzzling feature we observe in the monoclinic phase where the FWHM of $A_g$ (Te) mode increases while the width of the $A_g$ (Fe) modes decreases and returns back to the value it has in the tetragonal phase.

In the following we will discuss the relation of the Fe magnetic moment and anharmonicity. Solely chalcogen's (pnictogen's) antiphase vibrations that occur perpendicular to the Fe layer have an internal source of anharmonicity. Indeed, as one can see from Fig. 3, the $A_{1g}$ type of Te vibrations play the role of breathing-like mode for the chalcogen's (pnictogen's) tetrahedra. In a localized-electron framework some modulation of Fe-Te distances (i.e. radii of $Fe^{2+}$ - $3d^6$ ion) can induce a spin - state instability of $Fe^{2+}$ at which the intermediate spin state ($S = 1$) possesses the larger ionic radius and the low-spin state ($S = 0$) correspond to a smaller one (see, for instance, Refs. 34, 35). Such a modulation is specific for an intermediate spin state, i.e. for the Fe spin state with highest magnetic moment, because of its direct connection with an orbital reorder.

Furthermore, a few ground-state orbital ordering patterns which are consistent with the magnetically ordered structure have been addressed in iron pnictide superconductors [15]. Thus spin-orbital frustration accompanied with magnetic order with large Fe magnetic moments is connected to anharmonicity. Interestingly, as shown recently for FeO molecule [36], variations of the metal-ligand distances lead not only to a redistribution of the spin density but also to a modification of the adiabatic potential. And again, the $B_{1g}$ vibration of Fe ions should not be affected by this mechanism due to the Fe layer topology. We arrive at the conclusion that orbital degrees of freedom in general and particularly spin-orbital frustrations are responsible for the anomalous increase of width of the $A_g$ (Te) mode in FeTe compound. Note, that the fundamental role of the orbital degrees of freedom in the formation of electronic and magnetic excitations in FeSe$_{0.5}$Te$_{0.5}$ has been proven based on recent inelastic neutron studies [37].

By accepting these arguments one should check cases where this mechanism is not applicable, i.e. for the low value magnetic moment when the Fe$^{2+}$ is close to low spin state and orbital degrees of freedom become irrelevant. Indeed, a close inspection of available Raman data on $A_{1g}$ (As) phonons in undoped 1111 compounds [38] and 122 compounds [39, 40], all of which possess low value of magnetic moment [13], demonstrates that in the magnetically ordered state the width of this phonon mode always decreases under temperature lowering. That is in accordance with our expectation. One can also deduce that orbital fluctuations in 1111 and 122 materials, if they exist in paramagnetic phase, become suppressed in the magnetically ordered state.

The relevance of other mechanisms of electron-phonon coupling to the discussed phonons should be excluded by the following reason. There exist a remarkable resistivity drop in the magnetically ordered phase for both FeTe [11] and for SrFe$_2$As$_2$ [40], while the width of $A_{1g}$ (Te/As) modes demonstrate the opposite change with temperature. The only difference between the magnetic states of the FeAs/Te layers is the magnitude of iron magnetic moments and that point to the different involved orbital states.

The intensity data for the two phonons given in Fig. 2 (c, f) all exhibit an increase with decreasing temperature and does not show any anomaly within our experimental resolution.

We would like to highlight that we ascribe the giant splitting of originally degenerate $E_g$ modes in the monoclinic magnetic phase (see Table 1) to spin-lattice interaction of purely exchange (Coulomb) nature. The cases of symmetry reducing spin-lattice interaction of exchange nature are extremely rare. They can be very important because the respective distortion can be much stronger, compared to the regular case of relativistic spin-lattice interaction. Recall that normally, lattice distortions in magnetic materials are associated with relativistic spin-lattice interactions, which are usually much weaker.

To summarize all above arguments we stress that the $A_{1g}$ (chalcogen/pnictogen) phonon line width is a marker of the actual Fe orbital state in the parent compounds of iron based superconducturs. Orbital degrees of freedom should be taken into account in the systems with large magnetic moments.

**Acknowledgment**

This work was supported by Russian-Ukrainian grant No. 9-2010 and FRSF of Ukraine Grant No.F29.1/014 and the German Science Foundation via DFG LE 967/6-1 and SPP 1458. The calculations have been performed at the grid-cluster of Donetsk PhysTech NASU under support of grant N232. Yu.P. acknowledges partial support from the Swiss National Science Foundation (grant SNSF IZKOZ2_134161).

TABLE I. The results of phonon-mode calculations and a comparison with data form Raman experiments.

| Symmetry ($P4/nmm$) | Symmetry ($P2_1/m$) | Atoms and its displacements | Frequencies, cm$^{-1}$ | |
|---|---|---|---|---|
| | | | Theory ($P2_1/m$) Γ-point k=0 | Experimental data (80 K) |
| $E_g$ | $B_g$ | (Fe+Te, $u_{x+/-y}$) | 70.96 | - |
| $E_g$ | $A_g$ | | 78.60 | - |
| $A_{1g}$ | $A_g$ | (Te, $u_{1z}-u_{2z}$) | **152.9** | 155.2 |
| $E_u$ | $A_u$ | | 174.4 | - |
| $E_u$ | $B_u$ | (Fe+Te) | 176.6 | - |
| $B_{1g}$ | $A_g$ | (Fe, $u_{1z}-u_{2z}$) | **196.3** | 201.4 |
| $E_g$ | $B_g$ | (Fe+Te, $u_{x+/-y}$) | 208.1 | - |
| $E_g$ | $A_g$ | | 218.2 | - |
| $A_{2u}$ | $A_u$ | (Fe, $u_{1z}+u_{2z}$) | 247.1 | - |

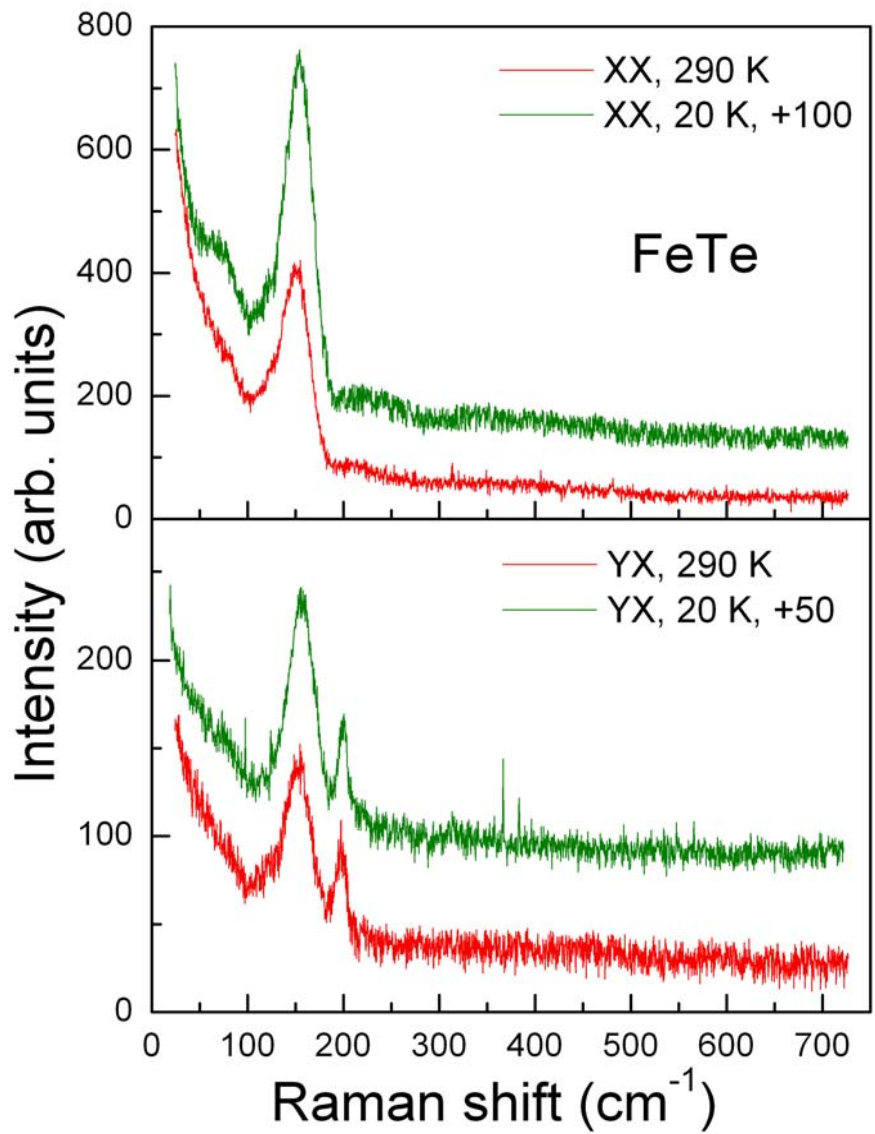

Fig. 1: Raman spectra of single crystal Fe$_{1.051}$Te taken in quasi-backscattering from the *ab*-plane at two temperatures. For clarity, the green curves are shifted in the vertical direction as indicated.

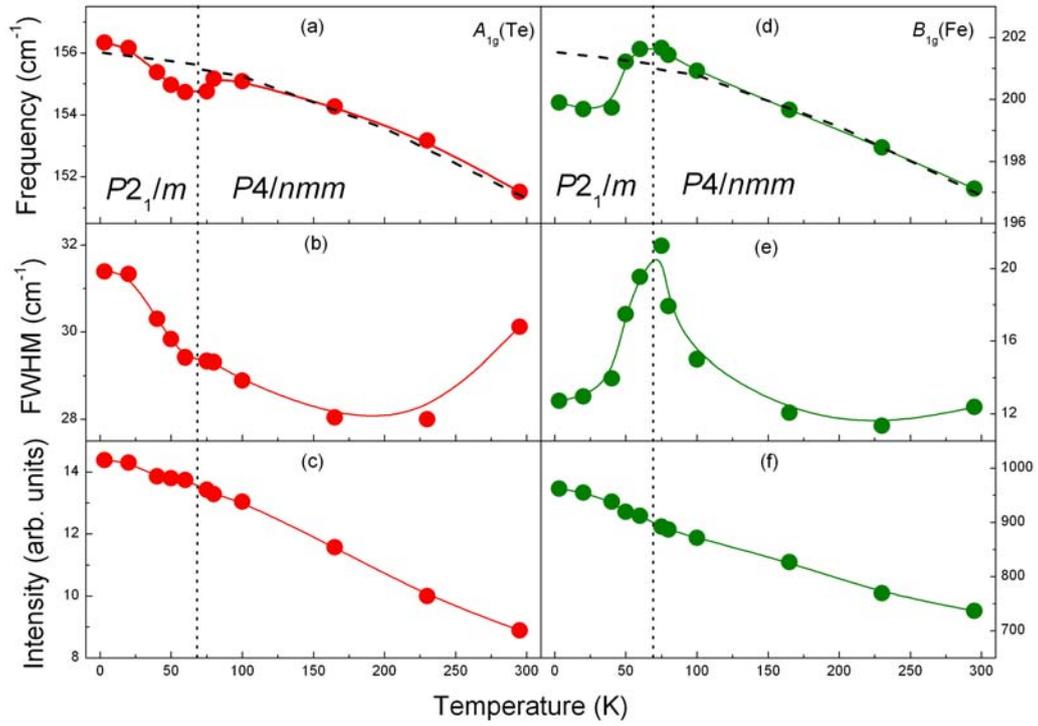

Fig. 2: Parameters of two phonon modes in $Fe_{1.051}Te$. (a, d) Temperature dependence of the frequency (solid circles) together with the fit using Eq. (1) (dashed lines); (b, e) Linewidth, FWHM; (c, f) integrated intensity. Solid lines in (b, c, e, f) are a guide to the eyes.

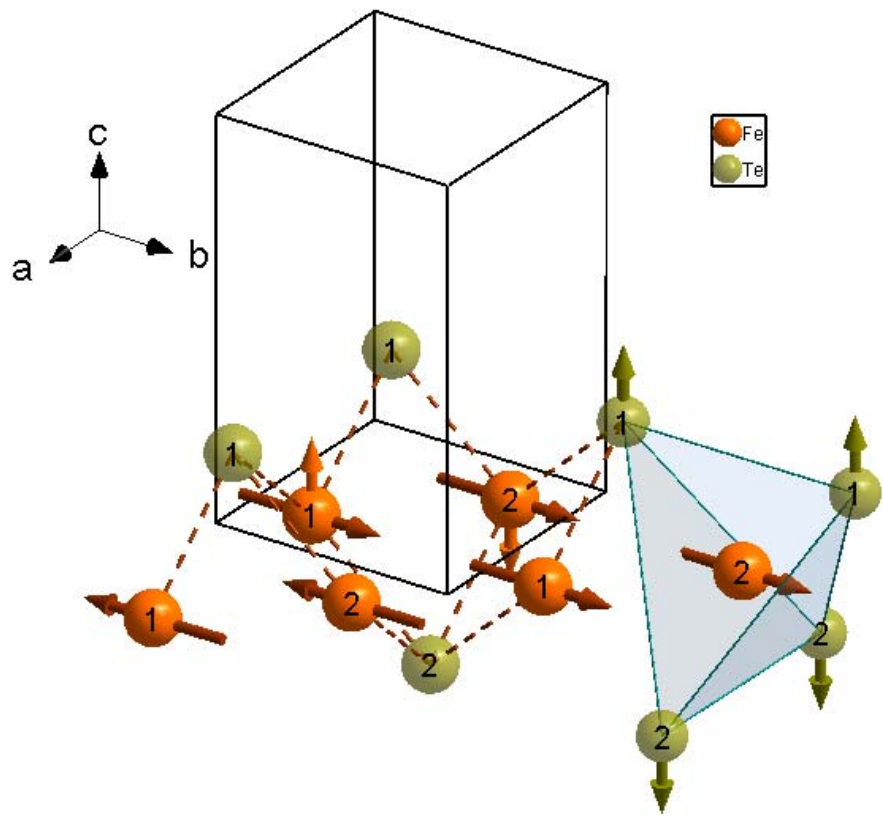

Fig. 3: Magnetic structure of FeTe in the monoclinic phase $P2_1/m$ with the magnetic propagation vector $k_1 = (\pi/a, 0)$. Magnetic moments are shown by dark arrows drawn through the atoms. $A_{1g}$ (Te) and $B_{1g}$ (Fe) ion's displacements are shown by arrows drawn from the ions.